\documentclass[twocolumn]{revtex4}
\usepackage{amsmath}
\usepackage{amssymb}
\usepackage{color}
\usepackage{ulem}
\usepackage{graphicx}
\newcommand{\rrangle}{\rangle\!\rangle}
\newcommand{\llangle}{\langle\!\langle}
\newcommand{\ra}{\rangle}
\newcommand{\la}{\langle}

\begin{document}

\title{Projection Hypothesis from the von Neumann-type Interaction with a Bose--Einstein Condensate}

\author{Eiji Konishi}
\email{konishi.eiji.27c@kyoto-u.jp}
\address{Graduate School of Human and Environmental Studies, Kyoto University, Kyoto 606-8501, Japan}

\date{\today}

\begin{abstract}
We derive the projection hypothesis in projective quantum measurement by restricting the set of observables.
This projection hypothesis accompanies a bipartite system with the von Neumann-type interaction, which consists of a quantum mechanical system, with a meter variable to be measured, and a quantum field theoretically macroscopic extended object, that is, a spatiotemporally inhomogeneous Bose--Einstein condensate in quantum field theory with the quantum coordinate, that is, the zero-energy Goldstone mode(s) of the spontaneously broken global spatial translational symmetry.
The key steps in the derivation are the return of the symmetry translation of this quantum coordinate to the inverse translation of the c-number spatial coordinate in quantum field theory and the reduction of quantum fluctuations to classical fluctuations with respect to the Goldstone mode(s) due to a superselection rule.
\end{abstract}

\maketitle

{\bf 1. Introduction.} $-$
Since the advent of the Copenhagen interpretation of quantum mechanics \cite{Bohr,Bohr2}, it has been a great challenge to resolve the measurement problem and the origin of the quantum state reduction that generates non-unitary time evolution of state vectors \cite{Neumann,London,Luders,Wigner,Everett,Decoherence,Decoherence2,Decoherence22,Decoherence3,Decoherence23,GRW,GRW2,MN1,MN2,Araki,Karolyhazy,Diosi,Diosi2,Penrose,4model}.

The {\it Copenhagen interpretation} of quantum mechanics provides two rules \cite{dEspagnat,NC}: I) the Schr${\ddot{\rm o}}$dinger equation for the state vector of an isolated quantum system, and II) the Born rule for the probability of obtaining a given measurement outcome, that is, an eigenvalue of a discrete observable to be measured at a projective quantum measurement performed on this system.
Notably, the time evolution generated by the Schr${\ddot{\rm o}}$dinger equation is unitary and linear, and the projective quantum measurement is a non-unitary process.

The {\it measurement problem} is a logical contradiction in this interpretation \cite{Neumann}.
Namely, when we consider the combined system of an isolated measured system and the measurement apparatus, no projective quantum measurement can be straightforwardly performed on that system.
This problem arises because the combined system is also an isolated system, so rule I) of the Copenhagen interpretation must be applied to it.
Therefore, one requires a second measurement apparatus for measuring the combined system, but according to the rules of the Copenhagen interpretation this regression continues infinitely.
To resolve this problem, von Neumann introduced the {\it projection hypothesis} as a terminal measurement apparatus in this infinite regression \cite{Neumann}.

We would like to reformulate projective quantum measurements and the measurement problem from modern perspectives.
The most important idea for this reformulation is to reinterpret the {\it probability} that appears in rule II) of the Copenhagen interpretation as the {\it statistical weight} that appears in the ensemble interpretation \cite{dEspagnat}.

In the ensemble interpretation of quantum mechanics, which is a framework for modern quantum measurement theory \cite{dEspagnat}, we consider an ensemble of copies of a quantum system in quantum {\it pure} states, which are given by state vectors.
In this language, a {\it pure} ensemble consists of copies of a state vector, and a non-trivial {\it mixed} ensemble consists of copies of distinct state vectors.
In this interpretation, we decompose the projective quantum measurement into two processes: a {\it non-selective measurement} process (also called a {\it decoherence process} or a {\it quantum-to-classical transition} \cite{ZurekPT}) and its subsequent {\it event-reading} process.
The non-selective measurement process is a dynamical process that makes the quantum pure ensemble of a given state vector of a quantum system be equivalent to the {\it classical} mixed ensemble (where {\it classical} means the absence of quantum interference in the system for a given eigenbasis) of eigenstates of a discrete measured observable whose statistical weights in the ensemble of copies of the system are given by the Born rule.\footnote{The term {\it non-selective measurement} intuitively means that, after only this dynamical process, events are in a classical mixed ensemble and no particular event is yet selected.}
The event-reading process is an informatical process that makes the classical mixed ensemble obtained by a non-selective measurement be the classical pure ensemble of an eigenstate of the measured observable.

In this reformulation, many proposals have been put forward for the physical realization of a non-selective measurement process (i.e., a decoherence process) \cite{Decoherence,Decoherence2,Decoherence22,Decoherence3,Decoherence23,GRW,GRW2,MN1,MN2,Araki,Diosi,Diosi2,4model}.
However, no proposal has been given for the physical realization of a subsequent event-reading process.

Significantly, the distinction between the projective quantum measurement \cite{NC} and the decoherence process \cite{Decoherence,Decoherence2,Decoherence22,Decoherence3,Decoherence23,GRW,GRW2,MN1,MN2,Araki,Diosi,Diosi2,4model} is the physical realizability of the event-reading process, that is, the final step of state reduction in the former.
For this reason, there has been no previous derivation of projective quantum measurement as an entirely physical process in the literature.

In this article, we consider a bipartite-interacting system $\Psi$, which consists of a {\it quantum mechanical} system $\psi$, with a meter variable ${\mathfrak M}$ (the pointer position or the energy of $\psi$), and a quantum field theoretically macroscopic extended object $A$ as a solitonic condensate in quantum field theory \cite{Umezawa1,Umezawa2,Umezawa3,Umezawa} with the quantum coordinate, and show that the event-reading measurement process of the meter variable ${\mathfrak M}$ accompanies this system as a physical process of the {\it state vector} $|\psi\ra$ of the system $\psi$ without modifying any law of quantum mechanics (e.g., the Schr$\ddot{{\rm o}}$dinger equation) except for the restriction of the set of observables of $A$ by a {\it superselection rule} (i.e., selection of the observables which commute with a given superselection charge) in $A$.
In other words, we derive the projection hypothesis in projective quantum measurement \cite{Neumann,London,Luders} as a physical substance.
For simplicity, we consider a model in one spatial dimension.
In our notation, hatted variables are quantum mechanical operators, and a superscript dot denotes the time derivative.

\begin{figure}[htbp]
\begin{center}
\includegraphics[width=0.32\hsize,bb=2 2 257 257]{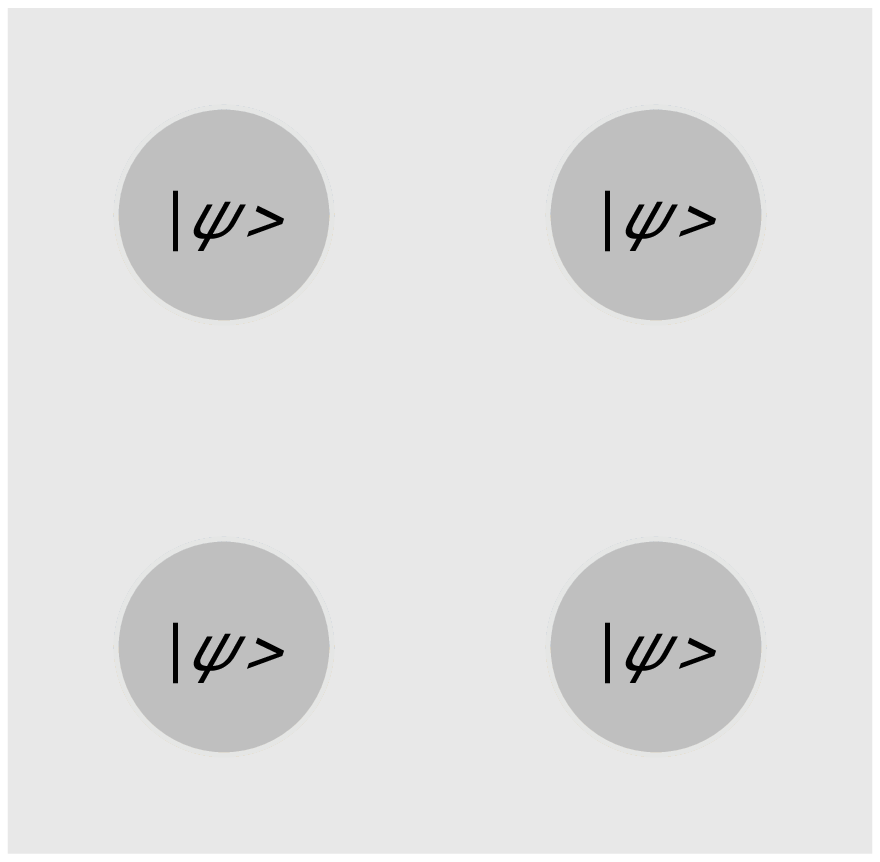}
\includegraphics[width=0.32\hsize,bb=2 2 257 257]{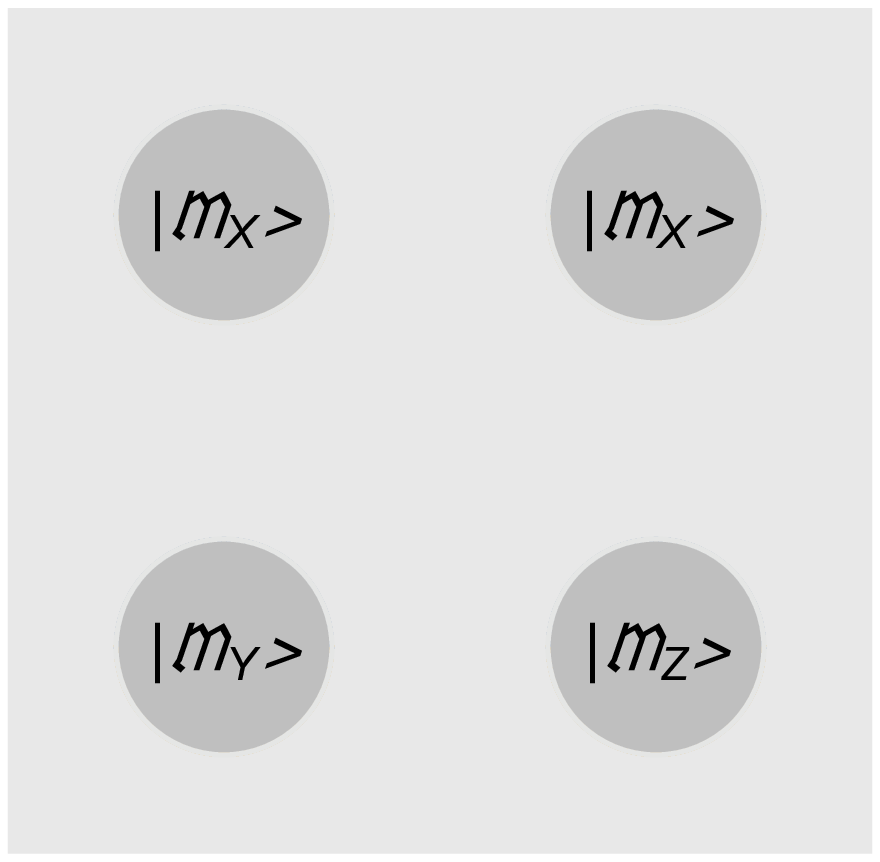}
\includegraphics[width=0.32\hsize,bb=2 2 257 257]{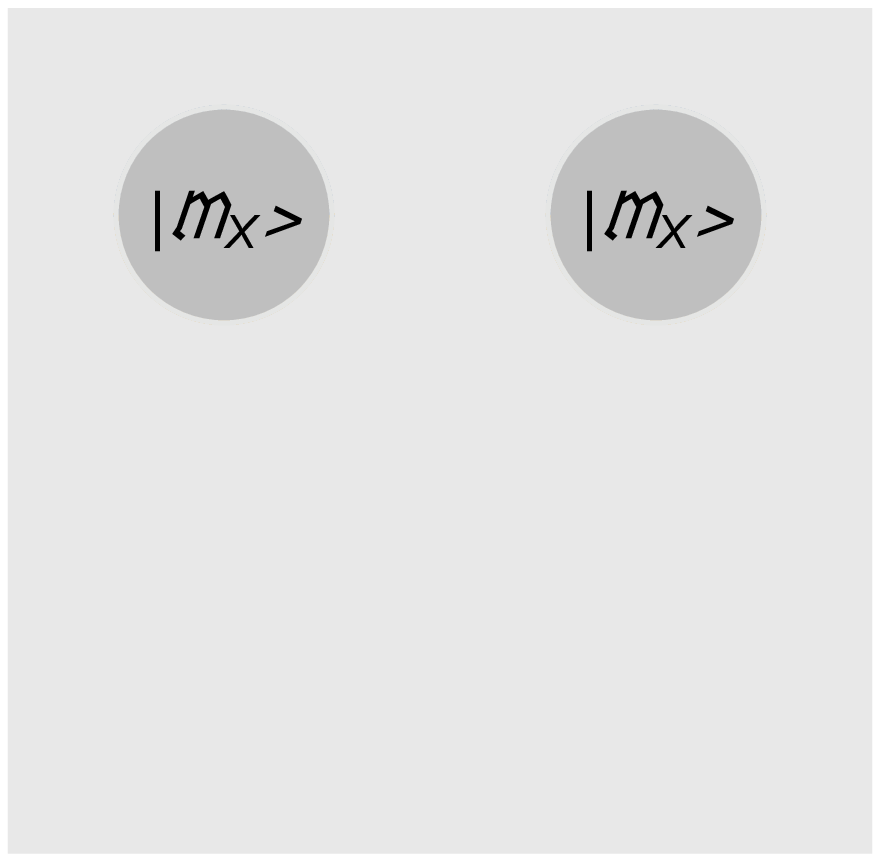}
\end{center}
\caption{Schematics of changes in the ensemble of four copies of the system $\psi$ during the {\it projective quantum measurement} process of the meter variable $\widehat{{\mathfrak{M}}}$.
The left panel shows the quantum pure ensemble of $\psi$ with the initial state vector $|\psi\ra=|{\mathfrak{M}}_x\ra/\sqrt{2}+|{\mathfrak{M}}_y\ra/2+|{\mathfrak{M}}_z\ra/2$ for three eigenstates $|{\mathfrak{M}}_x\ra$, $|{\mathfrak{M}}_y\ra$, and $|{\mathfrak{M}}_z\ra$ of $\widehat{{\mathfrak M}}$ with distinct eigenvalues.
The middle panel shows the classical mixed ensemble of $\psi$ after {\it non-selective ${\mathfrak M}$-measurement}.
The right panel shows the classical pure ensemble of $\psi$ after an {\it ${\mathfrak M}$-event reading} with the measurement event ${\mathfrak{M}}_x$.
The probability of obtaining this event is $1/2$.}
\end{figure}

In Fig. 1, we schematically show the two measurement processes of the meter variable $\widehat{{\mathfrak{M}}}$ in its projective quantum measurement applied to the quantum pure ensemble of an initial quantum pure state $|\psi\ra$ of the system $\psi$.

The rest of this article is organized as follows.
In Sec. 2, we describe the concept of a quantum coordinate, introduce the Hamiltonian of the model, and make three general remarks on quantum measurement theory and three remarks specific to our model.
In Sec. 3, we study the two measurement processes in our model.
Specifically, we show the mechanism and condition for the event-reading process and suggest a scheme for projective quantum measurement and an alternative.
In Sec. 4, we conclude this article.

\smallskip
\smallskip

{\bf 2. The model.} $-$ Before we introduce the Hamiltonian of the bipartite system $\Psi$, we describe our core concept, that is, the {\it quantum coordinate} of an extended object \cite{Umezawa}.

Once an extended object is created as a spatiotemporally inhomogeneous Bose--Einstein condensate in quantum field theory, global spatial translational symmetry breaks spontaneously due to the existence of this spatially inhomogeneous extended object in the ground state.
Then, there must exist a Goldstone mode, that is, the quantum coordinate (the center-of-mass position operator of the extended object) and its canonical conjugate as the zero-energy mode of the extended object (meaning it can move with no energy cost) to restore this spontaneously broken global symmetry.
This Goldstone mode restores the lost spatial translational invariance of the ground state by converting translation of the c-number spatial coordinate in quantum field theory into the inverse translation of the q-number quantum coordinate, where the quantum mechanical system of the Goldstone mode is spatially translationally symmetric.
Note that, although spontaneous symmetry breaking is a phenomenon in quantum field theory, the quantum coordinate is a {\it quantum mechanical} degree of freedom of the extended object, and no particle mode (like a phonon in a solid crystal) accompanies this quantum coordinate: the latter fact is because our Goldstone mode is the zero-energy mode and reflects that the center-of-mass position of the extended object is arbitrary, whereas the phonon modes, as gapless excitation modes, in a solid crystal are attributed to lattice vibration.

For the canonical conjugate, $\widehat{P}^x$, of this quantum coordinate, $\widehat{Q}^x$, of extended object $A$, the Hamiltonian $\widehat{H}_A$ of $A$ is given by \cite{Umezawa,Umezawa4}
\begin{equation}
\widehat{H}_A=\sqrt{\widehat{P}^2+\widehat{M}^2}\;,\label{eq:HA}
\end{equation}
where
\begin{equation}
\widehat{M}=M_0+\widehat{H}_0
\end{equation}
is the q-number rest energy of $A$ for the c-number mass $M_0$ of $A$ and the renormalized free Hamiltonian $\widehat{H}_0$ of the quasi-particles.
Here, the {\it quasi-particle} means that the ground state of $A$ is a vacuum state with respect to the quasi-particle, and the Fock space consists of quasi-particle pure states.
Note that (\ref{eq:HA}) is a relation among the matrix elements of $\widehat{H}_A$ in the extended Fock space of quasi-particles and the zero-energy mode \cite{Umezawa,Umezawa4}, and the rest energy $\widehat{M}$ of $A$ contains neither $\widehat{Q}^x$ nor $\widehat{P}^x$ but only the number operators of quasi-particles in $\widehat{H}_0$.

\begin{figure}[htbp]
\begin{center}
\includegraphics[width=0.8 \hsize,bb=4 2 260 173]{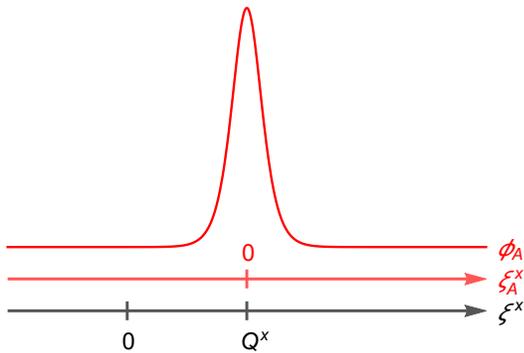}
\end{center}
\caption{Schematic of the order parameter $\phi_A=\phi_A(\xi_A^x,t)$ of extended object $A$ as a soliton and the spatial coordinate systems $\xi^x$ and $\xi_A^x$.
$\xi^x$ is the c-number spatial coordinate system in quantum field theory, and $\xi_A^x$ is the rearranged spatial coordinate system.
Note that $\xi^x$ and $\xi_A^x$ do not exist in quantum mechanics, in the sense that any operator in quantum mechanics consists of q-number degrees of freedom only.}
\end{figure}

Now, the Hamiltonian of the bipartite system $\Psi$ is given by
\begin{equation}
\widehat{H}_\Psi=\widehat{H}_\psi+\widehat{H}_A+\widehat{H}_{\psi{\mbox -}A}\;,
\end{equation}
where $\widehat{H}_\psi$ is the free Hamiltonian of $\psi$, and we introduce the von Neumann-type interaction Hamiltonian \cite{Neumann}
\begin{equation}
\widehat{H}_{\psi{\mbox -}A}=-\Lambda \widehat{{\mathfrak{M}}}\dot{\widehat{Q}^x}\ \ {\rm for}\ \ \dot{\widehat{Q}^x}=-\frac{i}{\hbar}[\widehat{Q}^x,\widehat{H}_A]\;,\label{eq:vN}
\end{equation}
which is spatially translationally invariant with respect to $A$.
In (\ref{eq:vN}), we adopt the interaction picture of $\Psi$ with respect to the free part $\widehat{H}_\psi+\widehat{H}_A$ of $\widehat{H}_\Psi$, and temporal linearity of the quantum coordinate
\begin{equation}
\widehat{Q}^x=\widehat{Q}_{\rm Sch}^x+\dot{\widehat{Q}^x}t
\end{equation}
holds for the Schr$\ddot{{\rm o}}$dinger picture quantum coordinate $\widehat{Q}_{\rm Sch}^x$ and a time parameter $t$ \cite{Umezawa2,Umezawa3,Umezawa}.
Then, in terms of eigenvalues, we obtain
\begin{eqnarray}
E_{\psi{\mbox -}A}t&=&-\Lambda {\mathfrak M}\dot{Q}^x t\\
&=&-{\mathfrak M}\delta_t Q^x\;,\ \ \Lambda\equiv 1\;,\label{eq:Eq1}
\end{eqnarray}
where we assume that $\Lambda$ is sufficiently strong to neglect the kinetic energies of $\psi$ and $A$.
Here, a symmetry translation of the quantum coordinate
\begin{eqnarray}
Q^x&\to&Q^x+\delta_t Q^x\;,\label{eq:Qx0}\\
\xi^x&\to&\xi^x\;,\label{eq:Nx0}
\end{eqnarray}
where $\xi^x$ is the c-number spatial coordinate system,
can be returned to the inverse translation of $\xi^x$:
\begin{eqnarray}
Q^x&\to&Q^x\;,\label{eq:Eq2}\\
\xi^x&\to&\xi^x+\delta_t \xi^x\label{eq:Nx}
\end{eqnarray}
in the space- and time-dependent order parameter (i.e., the vacuum expectation value of the desired boson Heisenberg field) $\phi_A$ of extended object $A$, where $\delta_t \xi^x$ is given by
\begin{eqnarray}
\delta_t \xi^x&=&\delta_t \xi_A^x\\
&=&-\delta_t Q^x\label{eq:Eq3}
\end{eqnarray}
for the rearranged spatial coordinate system (see Fig. 2)
\begin{equation}
\xi_A^x=\xi^x-Q^x\label{eq:xiAx}
\end{equation}
of $A$ satisfying
\begin{equation}
\phi_A=\phi_A(\xi_A^x,t)\;,
\end{equation}
which was derived in refs.\cite{Umezawa1,Umezawa2,Umezawa3,Umezawa}, subject to the non-relativistic-velocity condition $\dot{Q}^x\ll c$ \cite{Umezawa3,Umezawa}.\footnote{In (\ref{eq:xiAx}), our choice of sign on the quantum coordinate $Q^x$ follows ref.\cite{Umezawa}.}
This reversible return from (\ref{eq:Qx0}) and (\ref{eq:Nx0}) to (\ref{eq:Eq2}) and (\ref{eq:Nx}) in the order parameter $\phi_A$ of $A$ is from the Nambu--Goldstone theorem of quantum field theory and does not change the wave function of $A$
\begin{equation}
A(\xi^x=Q^x)=\la \xi^x=Q^x|A\ra\;,
\end{equation}
where the quantum mechanical sector of the zero-energy mode is used, but the quantum field theoretical sector of quasi-particles plays no important role for our purposes and is therefore ignored.

We make six remarks.

First, we make three general remarks on quantum measurement theory:
\begin{enumerate}
\item[i)] In quantum mechanics, the no-go theorem was proved by Wigner, Fine, and Shimony \cite{Wigner,Fine,Shimony} for the process of non-selective measurement in the combined system of the measured system and the measurement apparatus, so past proposals for this process introduced modifications of quantum mechanics (e.g., the restriction of the set of observables of a macroscopic measurement apparatus by a superselection rule in it; see refs.\cite{MN1,MN2,Araki}).

\item[ii)] The conversion process (i.e., the so-called {\it spectral decomposition} \cite{MN1,MN2}) of quantum eigenstates of a discrete microscopic measured observable into quantum eigenstates of a discrete macroscopic observable by using an apparatus such as the Stern--Gerlach apparatus is preliminary to and distinguished from the non-selective measurement process.
Here, the Stern--Gerlach apparatus (an applied spatially inhomogeneous magnetic field) converts the spin eigenstates of a half-spin particle with non-zero magnetic moment into macroscopically distinguishable quantum eigenstates of its spatial compartment position.
The spectral decomposition is a unitary process and is thus not involved in the measurement problem.

\item[iii)] The von Neumann-type interaction (\ref{eq:vN}) was originally used in conventional quantum measurement, which is realized by the trace out of the measurement apparatus, to entangle an initial quantum pure state of the measured system in the eigenbasis of the measured observable (corresponding to $\widehat{{\mathfrak M}}$) and the ready quantum state of the pointer position (corresponding to ${\widehat{Q}}^x$) of the measurement apparatus before the trace out of the measurement apparatus \cite{Neumann}.
Here, we emphasize that this conventional quantum measurement, in which information on the measurement apparatus is completely lost by the partial trace, is unsatisfactory in the context of the measurement problem and differs from the non-selective measurement in remark i).
\end{enumerate}

Second, the following three remarks are specific to our model:
\begin{enumerate}
\item[iv)] For a quantum mechanical position eigenvalue $q^x$ of a particle (not a condensate such as $A$), $\xi^x-q^x$ is not a rearranged spatial coordinate in quantum field theory: $\widehat{q}^x$ is not a quantum coordinate.
For this reason, our scheme for event reading would not work if the system $A$ is replaced with a different macroscopic system that is independent of spontaneous spatial translational symmetry breaking.

\item[v)] Spontaneous breakdown of the spatial translational symmetry of $A$ rearranges the Galilean transformation
\begin{equation}
\xi_{\rm old}^x\to \xi_{\rm old}^x+\dot{\xi}^xt
\end{equation}
to
\begin{equation}
\xi_{A,{\rm old}}^x\to \xi_{A,{\rm old}}^x+\dot{\xi}_A^xt\;,\label{eq:gauge}
\end{equation}
where the velocity $-\dot{\xi}_A^x$ relative to a fixed $\xi_{A,{\rm old}}^x$ is a {\it gauge}, in $\xi_A^x$, associated with time $t$ (i.e., arbitrariness in the coordinate transformation (\ref{eq:gauge}) in Galilean relativity).

\item[vi)] In our model, $\xi_A^x$ and $\dot{\xi}_A^x$ (relative to $\xi^x$) have simultaneous {\it classical} fluctuations, in a classical mixed state, which are the inverse of the simultaneous classical fluctuations of the quantum coordinate and its velocity, respectively.
Classicality (i.e., the absence of quantum interference) of these fluctuations is attributed to the orbital superselection rule in $A$ as a result of the macroscopicity of $A$ and will be shown by (\ref{eq:PsiE2}) in the next section.
\end{enumerate}


{\bf 3. Measurement processes.} $-$ In our picture, the time-evolution equation of the {\it state vector} $|\Psi(t)\rrangle$ of the bipartite system $\Psi$ with its interaction Hamiltonian (\ref{eq:vN}) (redefined as follows) is the Schr$\ddot{{\rm{o}}}$dinger equation.
We write it as
\begin{equation}
i\hbar\frac{d|\Psi(t)\rrangle}{dt}=\widehat{{\cal{H}}}_{\psi{\mbox -}A}|\Psi(t)\rrangle\;,\label{eq:Sch}
\end{equation}
where $|\Psi(t)\rrangle$ can be written as a direct sum \footnote{From (\ref{eq:dotQ}), the redefined velocity $\dot{{\cal Q}}^x$ is measurable simultaneously with the redefined quantum coordinate ${\cal Q}^x$.}
\begin{eqnarray}
|\Psi(t)\rrangle&=&\bigoplus_{\{({\cal Q}^x,\dot{{\cal Q}}^x)\}}|\psi(t,\xi^x={\cal Q}^x,\dot{{\cal Q}}^x)\rrangle \nonumber\\
&&A(\xi^x={\cal Q}^x,\dot{{\cal Q}}^x)\label{eq:Cl}
\end{eqnarray}
with $\{{\cal Q}^x\}$ (resp., $\{\dot{{\cal Q}}^x\}$) referring to the compartments of a partitioned original spectrum line with a constant width (resp., referring to those with the reciprocal constant width \cite{Neumann,dEspagnat}),
\begin{eqnarray}
|\psi(t,\xi^x={\cal Q}^x,\dot{{\cal Q}}^x)\rrangle\equiv|\psi(t)\ra |\xi^x={\cal Q}^x,\dot{{\cal Q}}^x\ra\;,
\end{eqnarray}
and the wave function of $A$
\begin{eqnarray}
A(\xi^x={\cal Q}^x,\dot{{\cal Q}}^x)=\la\xi^x={\cal Q}^x,\dot{{\cal Q}}^x|A\ra\;.
\end{eqnarray}
This direct sum structure (meaning the absence of quantum interference and classicality of fluctuations) with respect to the pair of the {\it redefined} quantum coordinate, ${\cal Q}^x$, and the {\it redefined} velocity, $\dot{{\cal Q}}^x$, arises from the macroscopicity of the extended object $A$ \cite{Araki,Ozawa}.
Namely, not fine and reciprocally fine measurement errors from the original $\widehat{Q}^x$ and $\dot{\widehat{Q}^x}$ accompany their redefinitions $\widehat{{\cal Q}}^x$ and $\widehat{\dot{{\cal Q}}^x}$, respectively, in the simultaneous eigenstates of $\widehat{{\cal Q}}^x$ and $\widehat{\dot{{\cal Q}}^x}$ \cite{Neumann,dEspagnat}.
Note that, within certain measurement errors, all {\it redefined} canonical observables of the quantum mechanical system that incorporate the measurement errors commute with each other (namely, a superselection charge exists as $\widehat{{\cal Q}}^x$ or $\widehat{\dot{{\cal Q}}^x}$), and the full set of simultaneous eigenfunctions of the redefined quantum coordinate operator and the redefined momentum operator forms a complete orthonormal system of the functional Hilbert space of this system: this is von Neumann's theorem in ref.\cite{Neumann}.
As a consequence, we obtained
\begin{equation}
[\widehat{{\cal Q}}^x,\widehat{\dot{{\cal Q}}^x}]=0\label{eq:dotQ}
\end{equation}
after this redefinition of the observables of the Goldstone mode of $A$ \cite{Neumann,dEspagnat} (i.e., after a restriction of the set of observables of $A$).
Using (\ref{eq:Cl}), the solution of the Schr$\ddot{{\rm{o}}}$dinger equation can be written as a direct sum \footnote{Here, the displacement $\delta_t{\cal Q}^x$ {\it in the phase} of the quantum state $|\Psi(t)\rrangle$ does not require the classical displacement $\delta_t{\cal Q}^x$ of extended object $A$.}
\begin{eqnarray}
|\Psi(t)\rrangle&=&\bigoplus_{\{({\cal Q}^x,\dot{{\cal Q}}^x)\}}\sum_nc_ne^{\frac{i\delta_t{\cal Q}^x}{\hbar}{\mathfrak{M}}_n}|{\mathfrak{M}}_n(\xi^x={\cal Q}^x,\dot{{\cal Q}}^x)\rrangle\nonumber\\
&&A(\xi^x={\cal Q}^x,\dot{{\cal Q}}^x)\label{eq:PsiE}
\end{eqnarray}
in the ${{{\mathfrak{M}}}}$-eigenbasis $\{|{\mathfrak M}\ra\}$ of the system $\psi$.
Using (\ref{eq:Eq3}), we rewrite (\ref{eq:PsiE}) as the easily understandable form
\begin{eqnarray}
|\Psi(t)\rrangle&=&\bigoplus_{\{(\Xi_A^x,\dot{\Xi}_A^x)\}}\sum_nc_ne^{-\frac{i\delta_t\Xi_A^x}{\hbar}{\mathfrak{M}}_n}|{\mathfrak{M}}_n(\Xi_A^x=0,\dot{\Xi}_A^x)\rrangle \nonumber\\
&&A(\Xi_A^x=0,\dot{\Xi}_A^x)\label{eq:PsiE2}
\end{eqnarray}
for the redefined and rearranged spatial coordinate system
\begin{equation}
\Xi_A^x=\xi^x-{\cal Q}^x
\end{equation}
of $A$ with the velocity $-\dot{\Xi}_A^x$ relative to $\xi^x$.
It is noteworthy that $\psi$ is {\it decoupled from} (i.e., in a product state with) $A$ in this mixture.
Equation (\ref{eq:PsiE2}) is equivalent to (\ref{eq:PsiE}), because their wave functions are the same.

Now, we examine a state change of $\psi$ with no dynamical element, which is attributed to {\it physical} equivalence of the time-dependent process of a state $|\psi(t)\ra$ in quantum mechanics with respect to the rearranged Galilean transformation
\begin{equation}
\Xi_{A,{\rm old}}^x\to \Xi_{A,{\rm old}}^x+(\dot{\Xi}_A^x-\dot{\Xi}_{A,{\rm old}}^x)t
\end{equation}
in quantum field theory (note Fig. 2 and the previous remark v)), as an informatical process.
In (\ref{eq:PsiE2}), we assume that the velocity $-\dot{\Xi}_A^x$, relative to $\xi^x$, (a gauge) has classical fluctuation (i.e., distinct classical events in the ensemble, ${\cal E}_A$, of copies of $A$) with the statistical weight of a classical event $\dot{\Xi}_A^x$
\begin{eqnarray}
w(\dot{\Xi}_A^x)=\sum_{\Xi_A^x}|A(\Xi_A^x=0,\dot{\Xi}_A^x)|^2\;.\label{eq:PSII}
\end{eqnarray}
For a time increment $\mu$ and two different classical events of increment $\delta_\mu\Xi_A^x(\equiv \dot{\Xi}_A^x\mu)$ in the ensemble ${\cal E}_A$, namely, two events
\begin{eqnarray}
\delta_\mu \Xi_{A,1}^x\neq \delta_\mu \Xi_{A,2}^x
\end{eqnarray}
that are attributed to the classical fluctuation of $\dot{\Xi}_A^x$, the equivalence $\simeq$ of two state vectors, which are evolved by the interaction Hamiltonian (\ref{eq:vN}) after redefinition of the observables of the Goldstone mode of $A$, up to an overall phase-factor difference \footnote{Here, the state of $A$ is traced out because it is decoupled from $|\psi(\mu)\ra$ in the mixture $|\Psi(\mu)\rrangle$.
Namely, we compare $|\psi_1\ra$ in $|\psi_1\ra|A_1\ra$ with $|\psi_2\ra$ in $|\psi_2\ra|A_2\ra$.}
\begin{eqnarray}
e^{-\frac{i\delta_\mu \Xi_{A,1}^x}{\hbar}\widehat{{\mathfrak{M}}}}|\psi(0)\ra\simeq 
e^{-\frac{i\delta_\mu \Xi_{A,2}^x}{\hbar}\widehat{{\mathfrak{M}}}}|\psi(0)\ra\label{eq:PSI}
\end{eqnarray}
holds in the (naively, due to the trace out of $A$, quantum mixed) ensemble, ${\cal E}_\psi$, of copies of $\psi$, which are obtained by reducing $\Psi$ in its copies, at $t=\mu$ if and only if the state vector $|\psi(0)\ra$ is an {\it eigenstate} of $\widehat{{\mathfrak{M}}}$ (namely, ${\cal E}_\psi$ is a classical pure ensemble at $t=0$).
Indeed, for a superposition
\begin{equation}
|\psi(0)\ra=c_1|{\mathfrak{M}}_1\ra+c_2|{\mathfrak{M}}_2\ra\;,\ \ {\mathfrak{M}}_1\neq {\mathfrak{M}}_2\;,
\end{equation}
there is discrepancy ratio
\begin{equation}
e^{-\frac{i}{\hbar}(\delta_\mu \Xi_{A,1}^x-\delta_\mu \Xi_{A,2}^x)({\mathfrak{M}}_1-{\mathfrak{M}}_2)}\not\equiv 1
\end{equation}
between the relative phases in the left- and right-hand sides of (\ref{eq:PSI}):
\begin{eqnarray}
&&e^{-\frac{i\delta_\mu \Xi_{A,1}^x}{\hbar}\widehat{{\mathfrak{M}}}}|\psi(0)\ra\nonumber\\
&&=c_1e^{-\frac{i \delta_\mu \Xi_{A,1}^x}{\hbar}{\mathfrak{M}}_1}|{\mathfrak{M}}_1\ra+c_2e^{-\frac{i \delta_\mu \Xi_{A,1}^x}{\hbar}{\mathfrak{M}}_2}|{\mathfrak{M}}_2\ra\;,\\
&&e^{-\frac{i\delta_\mu \Xi_{A,2}^x}{\hbar}\widehat{{\mathfrak{M}}}}|\psi(0)\ra\nonumber\\
&&=c_1e^{-\frac{i \delta_\mu \Xi_{A,2}^x}{\hbar}{\mathfrak{M}}_1}|{\mathfrak{M}}_1\ra+c_2e^{-\frac{i \delta_\mu \Xi_{A,2}^x}{\hbar}{\mathfrak{M}}_2}|{\mathfrak{M}}_2\ra\;.
\end{eqnarray}

When there exists quantum ${{\mathfrak{M}}}$-interference of $|\Psi(0)\rrangle$ with respect to the expectation values $\{\llangle \widehat{O}\rrangle\}$ of all observables $\{\widehat{O}\}$ of the system $\Psi$ selected by the orbital superselection rule in $A$ such that
\begin{equation}
\widehat{O}=\bigoplus_{\{(\Xi_A^x,\dot{\Xi}_A^x)\}} \widehat{O}(\Xi_A^x=0,\dot{\Xi}_A^x)\label{eq:Ost}
\end{equation}
holds \cite{Araki,Ozawa}, the classical fluctuation of events of increment $\delta_\mu\Xi_A^x$ cannot be described by using $|\psi(0)\ra$ or imposing (\ref{eq:PSI}) on $|\psi(0)\ra$ (instead, $\{\la \widehat{O}_\psi\ra\}$ is replaced with $\{{\llangle \widehat{O}\rrangle}\}$ such that
\begin{equation}
\widehat{O}(\Xi_A^x=0,\dot{\Xi}_A^x)=\widehat{O}_\psi\otimes|\Xi_A^x=0,\dot{\Xi}_A^x\ra\la \Xi_A^x=0,\dot{\Xi}_A^x|
\end{equation}
holds in (\ref{eq:Ost}), and $|\Psi\rrangle$ will be decohered with respect to ${\mathfrak{M}}$ by the von Neumann-type interaction $\widehat{H}_{\psi{\mbox -}A}$ as shown in refs.\cite{Araki,Ozawa,Konishi}).
This is because (\ref{eq:PSI}) is the condition for a state change of $\psi$ with no dynamical element (i.e., no change of time), and thus it cannot extinguish the quantum ${{\mathfrak{M}}}$-interference.

However, just after a non-selective ${\mathfrak M}$-measurement of a state vector $|\Psi\rrangle$ (a mixed state with respect to $A$), viewed as preparation of the state $|\Psi(-0)\rrangle$ just before the informatical process induced by (\ref{eq:PSI}) at $t=0$, quantum ${{\mathfrak{M}}}$-interference of $|\Psi(-0)\rrangle$ with respect to $\{{\llangle\widehat{O}\rrangle}\}$ cannot be observed in statistical data (namely, for $\widehat{O}$, its total interference is less than its uncertainty width $\Delta \widehat{O}$, with respect to the decohered state $|\Psi(-0)\rrangle$; see ref.\cite{Konishi2}):
\begin{eqnarray}
\bigl(\Delta \widehat{O}\bigr)^2&\equiv&\llangle \widehat{O}^2\rrangle -\llangle \widehat{O}\rrangle^2\\
&\gg& \Biggl(2\sum_{m<n}{\rm Re}\bigl[{\llangle \Psi_n|\widehat{O}|\Psi_m\rrangle}\bigr]\Biggr)^2\ {\rm for}\ {\rm all}\ \widehat{O}\;,
\end{eqnarray}
where $\{|\Psi_n\rrangle\}$ are the vector components of the state vector $|\Psi(-0)\rrangle$ in the ${\mathfrak M}$-eigenbasis.
Namely, this state $|\Psi(-0)\rrangle$ is {\it equivalent to} a classical mixed ensemble of ${\mathfrak M}$-eigenstates with respect to $\{{\llangle\widehat{O}\rrangle}\}$; the middle panel in Fig. 1 is thus realized.

At such a time, for {\it state vector} $|\psi(0)\ra$, (\ref{eq:PSI}) induces the informatical process, that is, a quantum state reduction that satisfies the Born rule as a result of the preceding non-selective ${\mathfrak M}$-measurement.
This is the subsequent ${\mathfrak M}$-event reading, which singles out an event of ${\mathfrak M}$ from the classical mixed ensemble; the right panel in Fig. 1 is thus realized.
This result achieves the aim of this article.

Based on this result, to realize projective quantum measurement of a discrete microscopic or spectrally decomposed observable, $\widehat{{\cal O}}$, of a quantum system $S_0$, we perform the following three steps as suggested in ref.\cite{Konishi}.
First, we perform the non-selective ${\cal O}$-measurement by using a macroscopic measurement apparatus $A^\prime$ based on the models in refs.\cite{Ozawa,Konishi}, where $\widehat{{\cal O}}$ is a discrete orbital observable, and ref.\cite{Araki2}, where $\widehat{{\cal O}}$ is a half spin.
Next, according to refs.\cite{Konishi,Konishi2}, we entangle the quantum state of the combined system $S=S_0+A^\prime$ in an ${\cal O}$-classical mixed ensemble with the quantum state of our measuring system $\Psi=\psi+A$ in the ${\mathfrak{M}}$-eigenbasis, starting from an ${\mathfrak M}$-eigenstate of $\Psi$.
Note that, by this entanglement, the non-selective ${\mathfrak M}$-measurement in the total system is in fact performed because the non-selective ${\cal O}$-measurement in $S$ is already performed.
Finally, we perform the ${\mathfrak M}$-event reading as shown above.

Alternatively, in this scheme for projective quantum measurement, the measured system $S_0$ with $\widehat{{\cal O}}$ (resp., the macroscopic measurement apparatus $A^\prime$) can be replaced with our system $\psi$ with $\widehat{{\mathfrak M}}$ (resp., our system $A$) when the spectrum of $\widehat{\dot{{\cal Q}}^x}$ of $A$ is sufficiently fine.

This is because, in this case, we can perform the non-selective ${\mathfrak M}$-measurement by using $A$ as a macroscopic measurement apparatus with the von Neumann-type interaction $\widehat{H}_{\psi{\mbox -}A}$ with $\psi$.
This result is shown in refs.\cite{Araki,Ozawa,Konishi,Konishi2}.

\smallskip
\smallskip

{\bf 4. Conclusion.} $-$
We conclude this article.
The assumptions in our derivation of the projection hypothesis in projective quantum measurement are spatiotemporally inhomogeneous Bose--Einstein condensation in quantum field theory and the orbital superselection rule in its quantum field theoretically macroscopic condensate.
This Bose--Einstein condensate can become part of our measuring system, which works as the projection hypothesis, in the presence of the von Neumann-type interaction with the other part, and the physical meaning of this condensate is that a boson Heisenberg field acquires a non-zero vacuum expectation value as the space- and time-dependent order parameter \cite{Umezawa}.
As a theoretical example in three spatial dimensions, the system of such a condensate can be realized as an equilibrium superradiant phase transition (i.e., photon condensation) in cavity quantum electrodynamics, where we treat the cavity wall (i.e., the boundary condition) as part of this system.
In this example, the boson Heisenberg field is a {\it spatially varying} cavity field coupled to the orbitals and spins of electrons in a magnetostatically unstable electronic system in the cavity in the thermodynamic limit, as reported in refs.\cite{PRL1,PRL2,PRB}; see also the no-go theorems for the equilibrium superradiant phase transition in the system of a spatially uniform cavity field coupled to matter proved in refs.\cite{Nature,PRB2}.
This universal scheme for generating the measuring systems gives substance to the projection hypothesis.

\end{document}